\documentclass[prl,twocolumn,showpacs,amsmath,amssymb,superscriptaddress]{revtex4}
\usepackage{graphicx}
\begin{document}
\title{Impurity-induced Local Density of States in a D-wave
Superconductor\\
Carrying a Supercurrent}
\author{Degang Zhang}
\author{C. S. Ting}
\affiliation{Texas Center for Superconductivity and Department of
Physics,\\
University of Houston, Houston, TX 77204, USA}
\author{C.-R. Hu}
\affiliation{Department of Physics, Texas A\& M University,
College Station, Texas, 77843, USA}
\begin{abstract}

The local density of states (LDOS) and its Fourier component induced by 
a unitary impurity in a supercurrent-carrying d-wave superconductor are 
investigated. Both of these 
quantities possess a reflection symmetry about the line passing 
through the impurity site and along the supercurrent if it is 
applied along the antinodal or nodal direction.  With increasing 
supercurrent, both the coherence and resonant peaks in the LDOS 
are suppressed and slightly broadened. Under a supercurrent 
along the antinodal direction, the coherence peaks split into 
double peaks. The modulation wavevectors associated with  
elastic scatterings of quasiparticles by the defect from one
constant-energy piece of the Fermi surface to another are
displayed as bright or dark spots in the Fourier space of the
LDOS image, and they may be suppressed or enhanced, and shifted
depending on the applied current and the bias voltage.

\end{abstract}
\maketitle

The understanding of the local physics in cuprate or high
temperature superconductors (HTS) is one of the most challenging
problems in condensed matter physics today.
Different from the conventional s-wave superconductors, the HTS
have very complex phase diagrams depending on doping and
chemical composition. It is also well established that the
superconducting order parameter in the cuprates has
predominantly d-wave symmetry [1]. The zero bias conductance
peak (ZBCP) in the tunneling spectroscopy of a normal
metal-cuprate superconductor junction with non-$(n0m)$ contact
provides one of the direct evidences for this symmetry [2]. Due
to the d-wave nature of the order parameter, impurities inserted
into cuprates can serve as an important tool to explore the
physics of HTS. Theoretical calculations of the local density of
states (LDOS) predicted that a strong potential scatterer could
induce a resonance peak near the Fermi level at sites near the
impurity [3,4]. This resonant peak near zero-bias voltage was
observed at and near the sites of Zn impurities in ${\rm
Bi}_2{\rm Sr}_2{\rm CaCu}_2{\rm O}_{8+\delta}$ by scanning
tunneling microscopy (STM)[5].  In addition, an interference
pattern with four-fold symmetry was also detected in the STM
image [5].

When a superconductor carries a supercurrent ($J_s$), Cooper
pairs with finite momentum appear in the system. This would
drastically affect the electronic structure of the
superconductor including the elementary excitation spectrum, the
order parameter symmetry, and the tunneling spectroscopy [6,7,8].
With increasing supercurrent velocity, the superconducting
order parameter can be depressed. Meanwhile, the supercurrent
density first increases monotonously and then arrives at a
maximum value, which  is called the critical current density.
Beyond that, superconductivity becomes unstable and collapses to
the normal state. So the supercurrent in the stable regime can be
also used as a probe to further understand the quasiparticle
excitations in HTS. Moreover, a better understanding of the
property of a superconductor under an applied $J_s$ may have the
potential for device applications.

In Ref. [7], we have studied the tunneling conductance
characteristics between a normal metal and a d-wave
superconductor (dSC) carrying a supercurrent parallel to the interface of 
the junction.  It was shown that
for sufficiently large applied current, the
midgap-surface-state-induced ZBCP splits into two peaks in the
tunneling regime. So far there exist no experimental measurements 
which could  be used to compare with our theoretical predictions. 
The closest tunneling experiment to the idea in Ref. [7] was done 
on YBCO under a spin injected current [9], it would be interesting 
to see that the similar experiment will be performed on an HTS 
sample carrying a supercurrent in the near future.  As a natural 
extension of our previous work [7], here we examine the LDOS induced 
by a strong defect which replaces a ${\rm Cu}^{2+}$ ion in the top 
CuO layer of a current carrying HTS. This strong defect could 
either be a unitary impurity like ${\rm Zn}^{2+}$ or simply a 
Cu-vacancy, and is well known to induce a near-zero-bias resonant 
peak (NZBRP) next to the site of the defect in a dSC without 
$J_s$ [3,4]. In the following, we investigate the LDOS images 
and their Fourier components for several values of the bias 
energy E and $J_s$. In addition the LDOS at sites next to and 
far away from  the impurity as a functions of E are calculated 
and their behaviors under various $J_s$ will be presented and 
discussed.  

The BCS Hamiltonian describing the impurity effects
in a dSC carrying a supercurrent can be written as
$$H=\sum _{{\bf k}\sigma}(\epsilon_{\bf k}-\mu)
c^+_{{\bf k}\sigma}c_{{\bf k}\sigma}+\sum_{\bf k}[
\Delta_{{\bf q}_s}({\bf k})c^+_{{\bf k}+{\bf q}_s\uparrow}
c^+_{-{\bf k}+{\bf q}_s\downarrow}+{\rm h.c}]$$
$$+V_s\sum_{\sigma}c^+_{0\sigma}c_{0\sigma},
\eqno{(1)}$$
where $\epsilon_{\bf k}$ is the band structure of
the d-wave superconductor, $\mu$ is the chemical
potential to be determined by doping,
$V_s$ is the on site potential of the
nonmagnetic impurity located at the center of lattice,
${\bf q}_s = (m*/2){\bf v}_s$ with ${\bf v}_s$ the
supercurrent velocity, and $m*$ the mass of a Cooper
pair, $\Delta_{{\bf q}_s}({\bf k}) =
\Delta_{{\bf q}_s} {\rm cos}(2\theta)$ is the
superconducting order parameter in the presence of $J_s$,
$\theta$ is the angle between the wave vector ${\bf k}$
and the antinodal direction of the d-wave superconductor, and
$\Delta_{{\bf q}_s}$ is determined by the gap equation [7]
$$\pi{\rm ln}\frac{\Delta_0}{\Delta^q}=
    \int_{\geq} d\theta {\rm cos}^2(2\theta)
     {\rm ln}[g(\phi)+\sqrt{g^2(\phi)-1}],
  \eqno{(2)}$$
where
$$g(\phi)\equiv\frac{2q}{\Delta^q}|\frac
{{\rm cos}(\theta-\phi)}{{\rm cos}(2\theta)}|,
~~~q\equiv\frac{q_s}{k_F},~~~
\Delta^q\equiv\frac{\Delta_{{\bf q}_s}}{E_F},
\eqno{(3)}$$
$k_F$ and $E_F$ are the Fermi momentum
and energy, respectively, $\phi$ is
the angle between ${\bf q}_s$ and the
antinodal direction, and the integraton in
Eq. (2) is from 0 to 2$\pi$ with the constraint
$g^2(\phi)-1\geq 0$. The solutions of Eq. (2)
with $\phi=0$ and $\frac{\pi}{4}$ are presented in Fig. 1(a).
In Ref. [7], we  also derived the thermodynamic
critical currents $j_{sc}(0)=0.238env_F\Delta_0$
at $q=q_c(0)=0.35\Delta_0$ and
$j_{sc}(\frac{\pi}{4})=0.225env_F\Delta_0$ at
$q=q_c(\frac{\pi}{4})=0.39\Delta_0$ for supercurrent $J_s$ along
the antinodal
and nodal directions, respectively [see Fig. 1(b)].
Similar results on the order parameter and the
critical currents have also been obtained in Ref. [8].

The Hamiltonian (1) is exactly soluble by the Bogoliubov 
transformation and the Green's function technique [10]. 
After a tedious but straightforward calculation, we obtain 
the expression for LDOS near a strong impurity in
the unitary limit (i.e. $V_s\rightarrow \infty$)

\begin{widetext}
$$\rho({\bf r},\omega)=\rho_0({\bf r},\omega)
+\delta \rho({\bf r},\omega),~~~~~
\rho_0({\bf r},\omega)=-\frac{2}{\pi{\cal N}}
  {\rm Im}\sum_{{\bf k},\nu}\xi^2_{{\bf k}\nu}
({\bf q}_s)G^0_{{\bf k}\nu}({\bf q}_s,i\omega_n)
|_{i\omega_n\rightarrow \omega+i0^+},$$
$$\delta \rho({\bf r},\omega)=-\frac{2}{\pi{\cal N}^2}
     {\rm Im}\sum_{{\bf k},{\bf k}^\prime}
      \sum_{\nu,\nu^\prime=0,1}D({\bf q}_s,i\omega_n){\rm cos}
      [({\bf k}-{\bf k}^\prime)\cdot {\bf r}]
 \{\xi^2_{{\bf k}\nu}({\bf q}_s)
          \xi^2_{{\bf k}^\prime\nu^\prime}({\bf q}_s)
   b({\bf q}_s,i\omega_n)
$$
$$ -2(-1)^\nu\xi_{{\bf k}\nu}({\bf q}_s)
          \xi_{{\bf k}\nu +1}({\bf q}_s)
          \xi^2_{{\bf k}^\prime\nu^\prime}({\bf q}_s)
       c({\bf q}_s,i\omega_n)
      +(-1)^{\nu+\nu^\prime}\xi_{{\bf k}\nu}({\bf q}_s)
       \xi_{{\bf k}\nu +1}({\bf q}_s)
      \xi_{{\bf k}^\prime\nu^\prime}({\bf q}_s)
      \xi_{{\bf k}^\prime\nu^\prime+1}({\bf q}_s)$$
$$  \times a({\bf q}_s,i\omega_n)
      \}G^0_{{\bf k}\nu}({\bf q}_s,i\omega_n)
       G^0_{{\bf k}^\prime\nu^\prime}({\bf q}_s,i\omega_n)
     |_{i\omega_n\rightarrow \omega+i0^+},\eqno{(4)}$$
\end{widetext}
where ${\cal N}$ is the site number of lattice, and
$$E_{{\bf q}_s}({\bf k})=\sqrt{[\frac{1}{2}
(\epsilon_{{\bf k}+{\bf q}_s}+\epsilon_{-{\bf k}+{\bf q}_s})
-\mu]^2+\Delta^2_{{\bf q}_s}({\bf k})},$$
$$\xi^2_{{\bf k}\nu}({\bf q}_s)=\frac{1}{2}
[1+(-1)^\nu\frac{\frac{1}{2}(\epsilon_{{\bf k}+{\bf q}_s}
+\epsilon_{-{\bf k}+{\bf q}_s})-\mu}{E_{{\bf q}_s}({\bf k})}],$$
$$\xi_{{\bf k}0}({\bf q}_s)\xi_{{\bf k}1}({\bf q}_s)
=\frac{\Delta_{{\bf q}_s}({\bf k})}{2E_{{\bf q}_s}({\bf k})},$$
$$G^0_{{\bf k}\nu}({\bf q}_s,i\omega_n)=
\frac{1}{i\omega_n-\frac{1}{2}(\epsilon_{{\bf k}+{\bf q}_s}
-\epsilon_{-{\bf k}+{\bf q}_s})-(-1)^\nu
E_{{\bf q}_s}({\bf k})},$$
$$a({\bf q}_s,i\omega_n)=\frac{1}{{\cal N}}\sum_{{\bf k},\nu}\xi^2_{{\bf
k}\nu}({\bf q}_s)G^0_{{\bf k}\nu}({\bf q}_s,i\omega_n),$$
$$b({\bf q}_s,i\omega_n)=\frac{1}{{\cal N}}\sum_{{\bf k},\nu}\xi^2_{{\bf
k}\nu+1}({\bf q}_s)G^0_{{\bf k}\nu}({\bf q}_s,i\omega_n),$$
$$c({\bf q}_s,i\omega_n)=\frac{1}{{\cal N}}\sum_{{\bf
k},\nu}(-1)^\nu\xi_{{\bf k}\nu}({\bf
q}_s)\xi_{{\bf k}\nu+1}({\bf q}_s)
G^0_{{\bf k}\nu}({\bf q}_s,i\omega_n),$$
$$D({\bf q}_s,i\omega_n)=\frac{1}{c^2({\bf q}_s,i\omega_n)
-a({\bf q}_s,i\omega_n)b({\bf q}_s,i\omega_n)}.\eqno{(5)}$$

Obviously, when a supercurrent is appplied, the quasiparticle
energy has a momentum-dependent shift
$\frac{1}{2}(\epsilon_{{\bf k}+{\bf q}_s} -\epsilon_{-{\bf
k}+{\bf q}_s})$ [see the bare Green's function $G^0_{{\bf
k}\nu}({\bf q}_s,i\omega_n)$ in Eq. (5)], which leads to
different gaps for different momentum directions of a
quasi-particle. This strongly modifies the LDOS and its Fourier
components patterns. We shall see that the variation of these
patterns is sensitive to the supercurrent applied and the bias
voltage. However, they always have a reflection symmetry at an
arbitrary energy if a supercurrent is applied along the
antinodal or nodal directions. From Eqs.(4) and (5), we
calculate the LDOS at several different energies and
supercurrent velocities with the defect located at the center of
a ${\cal N}=400\times 400$ lattice. Here we adopt the band
structure of ${\rm Bi}_2{\rm Sr}_2{\rm CaCu}_2{\rm
O}_{8+\delta}$ given by Norman {\it et al.} $\epsilon_{\bf
k}=-0.5951({\rm cos}k_x+{\rm cos}k_y)/2 +0.1636{\rm cos}k_x{\rm
cos}k_y -0.0519({\rm cos}2k_x+{\rm cos}2k_y)/2 -0.1117({\rm
cos}2k_x{\rm cos}k_y+{\rm cos}k_x{\rm cos}2k_y)/2 +0.0510{\rm
cos}2k_x{\rm cos}2k_y -\mu$ (eV) [11] , which corresponds to that
of free electrons with $k_F=1.639$ and $E_F=0.4203$ eV, for the
chemical potential $\mu=-0.1238$
eV for optimal doping (15$\%$). Choosing $\Delta_0=44$ meV, and
$\Delta_{\bf q}$ can be obtained from Fig. 1(a).

Fig. 2 shows the $20\times 20$ images of the LDOS $\rho({\bf r},\omega)$
at different energies and supercurrents with the impurity at its center.
In order to understand the image patterns in Fig. 2, we plot the schematic
Fermi surface of an optimally doped HTS in the first Brillouin zone as
shown in Fig. 3. In an STM experiment and when a quasiparticle is created
near the Fermi surface at point O, this quasiparticle may be scattered
elastically by the defect to other equivalent points (such as A, B, C, D,
E, F and G) near the Fermi surface [12-16]. The wavevectors connecting O
and the other points are
referred as the modulation wavevectors and they are labeled as $q_A$,
$q_B$, etc., up to $q_F$.
If the point O is at the middle of the Fermi curve in Fig. 3,
then the quasiparticle is at the nodal point. If O is
moved to the zone boundary, then the quasiparticle is at the
antinodal point. Because of the d-wave nature of the
superconductivity, little energy is required to create a
quasiparticle at the nodal point.  But to create a quasiparticle
at the antinodal point, a large bias energy in the order of the
superconductivity gap is needed in an STM experiment.

From Fig. 2, the LDOS at the impurity site vanishes regardless
of the bias energies and the strength of $J_s$, and it
has the strongest intensity at the sites next to the defect when
the quasi-particle energy (or the bias voltage times $e$)
$\omega = 0$ meV [see Fig. 2c(0)]. Near the impurity, the LDOS
has a pattern of 4-fold symmetry with energy-dependent
modulations in the absence of a supercurrent [Fig. 2a(0) to
2e(0)].

When $|\omega |$ = 0 and 16 meV, the resonant peaks still
show up at $(0,\pm 1)$ and $(\pm 1,0)$ and the modulation
with the periodicity $\sim 2a$ is along $45^0$ from the
Cu-O bonds. For $\omega$ = 0 meV, the point O is at the nodal
point, and the modulation in Fig. 2c(0) comes from the
wavevector $q_D = q_E$ in Fig. 3. For $|\omega |$ = 16 meV, the
pattern seems to be a result of combined contributions from
$q_B$ and other modulation wavevectors along the directions
$(\pm\pi, \pm\pi)$.

When $|\omega |$ = 44 meV, The point O moves to the zone
boundary or the antinodal point.  The modulation wavevector
$q_F = 2\pi$. This would give rise to  x- and y- oriented (or
Cu-O bond oriented)) stripe-like structure with the
periodicity $\sim a$ in the LDOS around the impurity, and this
can be seen clearly in Fig. 2a(0). The LDOS here also exhibits a
checkerboard pattern close to the impurity site due to the
combined effect of the x- and y-oriented stripes.

When a supercurrent $J_s$ is applid along the antinodal direction
(i.e. $\phi=0$ along the x-direction), the intensities of the
resonance peaks on points $(0,\pm 1)$ are higher than those
on the points $(\pm 1,0)$ at $|\omega | =$ 0 and 16 meV.
Near the critical current $j_{sc}(0)$, the LDOS developes a
modulation perpendicular to the direction of supercurrent at
$|\omega | =$ 16 meV. When $|\omega | =$ 44 meV, the intensity of
the modulation parallel to $J_s$ becomes
smaller than that perpendicular to $J_s$. When $J_s$ is applied
along the nodal direction (i.e $\phi=\pi/4$ from the x-axis),
the LDOS patterns only have minor changes except that some
brighter spots near the defect site appear at $|\omega | =$ 16
meV. We note that with increasing $J_s$, the maxima of
the LDOS at $|\omega | =$ 0 and 44 meV are suppressed while those
at $|\omega | =$ 16 meV are enhanced.

In order to further understand the supercurrent effects,
we also calculate the images for the Fourier component of the LDOS
(FCLDOS)( see Fig. 4).
It can be easily seen that the influence of the applied
suppercurent on the FCLDOS is more dramatic than on the LDOS
image. Here some modulation wavevectors corresponding to the
elastic scattering of quasiparticles from one point of the Fermi
surface to another point as shown in Fig. 3 can be clearly
identified in Fig. 4.

When $\omega =$ 0 meV and $q_s =$ 0, the modulation 
wavevectors $q_A = q_C = q_F$ and $q_D = q_E$ due to the 
nodal quasiparticle scattering are clearly seen in the 
FCLDOS patterns [Fig. 4c(0)]. We note that the dip at 
$q_{D,E}$ has a strong intensity, which causes the
LDOS to have a modulation along $45^0$ to Cu-O bonds
[Fig. 4c(0)]. At the critical currents $J_{sc}(0)$ and
$J_{sc}(\frac{\pi}{4})$, the dips are supressed, but their
positions seem not to shift [Fig. 4c(1) and 4c(2)].

However, at higher energy, the case becomes more complicated.
At $|\omega | =$ 16 meV, the dips at $q_A$ and $q_F$ are clearly
visible in the absence of supercurrent [Fig. 4b(0) and 4d(0)].
At four corners of the first Brillouin zone, there are four 
arcs due to the scatterings of the quasiparticles by the 
defect from one equal-energy banana countor (e.g. arcs OB in 
Fig.2) to the opposite contour (e.g. arc DE in Fig. 3). It is
these arcs that mainly produce the charge modulation along $45^0$
from the Cu-O bonds [Fig. 2b(0) and 2d(0)]. At $\omega = -16$
meV, we note that the peaks associated with $q_B$ are absent.
Instead, four new dips at $q_2$ show up. When a $J_s$ is
applied, these peaks and arcs for $|\omega |$=16 meV are
supressed or enhanced, and even vanish near the critical
current, but their position have little shift [see Fig. 4b(1),
4b(2), 4d(1) and 4d(2)].

When $|\omega | =$ 44 meV, Fig. 4a(0) and 4e(0) show the images
of the FCLDOS associated with the scattering of antinodal
quasiparticles. The peaks correspond to $q_A$ and $q_1$ are
clearly seen here. The modulation wavevectors $q_1$ are due to
the superposition of those peak arcs induced by the scatterings
of quasiparticles from one antinodal part of banana contour to
the neighboring part. Obviously, the equal-energy banana contour
becomes wide for $J_s$ along the antinodal direction while it
shrinks to a point for $J_s$ along the nodal direction [Figs.
4a(1), 4a(2), 4e(1) and 4e(2)]. From Fig. 2 and Fig. 4, it can
be seen clearly that the effect of $J_s$ on the FCLDOS is much
pronounced than that on LDOS.

In order to examine the change of modulation wavevectors with
$J_s$, we present the FCLDOS along the antinodal and nodal
directions at $\omega = 0$ meV and for several supercurrent
strengths in Fig. 5. With increasing the supercurrent velocity,
the dips associated with $q_{A,C,F}$ are suppressed and finally
disappear [see Fig. 5(a) and 5(c)]. However, the dips
corresponding to $q_{D,E}$ are only suppressed for $J_s$ along
the antinodal direction or $\phi = 0$ while they are first
enhanced, then weakened, and have a tiny shift for the $J_s$
along the nodal direction or $\phi = \frac{\pi}{4}$ [Fig. 5(b)
and 5(d)]. Similar results hold at higher energy. We note that
the dips at $|q| = 1.6\pi$ in Fig. 5(a) and 5(c) cannot be
induced by quasiparticle scatterings, which are also suppressed
with increasing $J_s$. We think that these dips at $q_{A,C,F}$
and $q_{D,E}$ are due to the manifestation of quasiparticle
destructive interference due to the sign change of the $d$-wave
gap function on the Fermi surface.

We have obtained the LDOS and FCLDOS induced by a strong defect
such as a Zn impurity. Now we turn our attention to the STM
experiments. In the STM experiments [5], a zero bias resonant
peak was observed at the Zn sites. However, theoretical
calculations give a vanishing LDOS at the impurity sites,
contrary to the experimental observation (see Fig. 2). Because
of a Bi atom in the top (BiO) layer and, more importantly, an O
atom in the second (SrO) layer block the tunneling current
coming from the STM tip to directly probe the impurity site
[17], the experimentally observed LDOS at the impurity or Cu
site should be approximately equal to the sum of those on four
nearest neighbor sites around it [18], i.e.
$$\rho_{\rm expt}({\bf R},\omega)\approx \sum_{\bf \delta}
\rho ({\bf R}+{\bf \delta},\omega), \eqno{(6)}$$
where ${\bf \delta}$ denote the nearest neighbor sites of
the impurity or Cu ions.

Taking into account this blocking effect, we present
$\rho_{\rm expt}({\bf R}, \omega)$ curves at the impurity site
$(0,0)$ and the points $(0,-1)$, $(1,-1)$ and $(2,0)$ for
several $J_s$ along $\phi = 0$ and $\phi = \pi/4$, respectively
in Fig. 6. Obviously, the NZBRP on the impurity site and its
neighbor sites are strongly suppressed and only slightly
broadened with increasing $J_s$ in both directions. No splitting
is clearly visible. On the other hand, the superconducting
coherence peaks shows some suppression and splitting near the
critical current along the antinodal direction with $\phi = 0$
while their separation widens with increasing $J_s$ along the
nodal direction with $\phi = \pi/4$. We further notice that the
suppression of the NZBRP is insensitive to the direction of $J_s$.
A relevant work [20] studied the NZBRP due to a unitary impurity in the 
presence of a magnetic field and away from the vortex cores. The magnetic 
field effect was considered by including the Doppler shifts [21] in the 
energies of the quasiparticles and it thus generates circulating 
supercurrent in the sample which is similar but not identical to the case 
we studied. The NZBRP displayed in Fig. 1 of Ref. [20] and Fig. 6 in the 
present paper are both suppressed by the magnetic field B or $J_s$, but it 
appears that their LDOS due to the impurity seem to lose a lot of  
spectral weight while ours practically remains as a constant as B or $J_s$ 
increases.  Further work is needed in order to understand the difference 
between these two works.     
 
In summary, we have investigated the supercurrent effects on the
impurity resonance states in d-wave superconductors. The LDOS and 
FCLDOS patterns induced by a strong impurity have a reflection symmetry 
if a supercurrent is applied along the antinodal or nodal directions. 
The suppression and broadening of the resonant peak and the 
superconducting coherence peaks are due to the anisotropic gap induced by 
a supercurrent. Future STM experiments need to be performed in order to 
test these predictions. On the other hand, the 
midgap-surface-state-induced ZBCP in the tunneling conductance 
characteristics between a normal metal and a d-wave
superconductor (dSC)[7] looks similar to the strong impurity induced 
NZBRP. 
But their dependences on the applied supercurrent $J_s$ are quite 
different. 

When  strong defects such as Cu vacancies and  micro-crystals with edges 
exposed to (110) direction are both present on the surface of a HTS 
sample, STM experiments should be able to distinguish them by analyzing 
the $J_s$ dependences of the zero-bias conductance peak induced by an 
isolated defect and that induced by the surface midgap states.

The authors wish to thank Prof. S. H. Pan for helpful Discussions.  
This work was supported by the Texas Center for Superconductivity and 
Advanced Materials at the University of Houston and by the Robert A. 
Welch Foundation (Ting).

Fig. 1: Dependences of the superconducting order parameter
on the normalized supercurrent-velocity parameter
$q$ for a d-wave superconductor (a) and the corresponding
dependences of supercurrent density on $q$ (b).

Fig. 2: The $20\times 20$ images of the LDOS $\rho({\bf r},\omega)$ at
$\omega=-44$ meV, $-16$ meV, 0 meV, 16 meV and 44 meV (from top to bottom) 
and $q=0, q_c(0)$ and $q_c(\frac{\pi}{4})$
(from left to right) for a unitary impurity at its center. 

Fig. 3: Schematic Fermi surface of high-$T_C$ cuprate superconductor.
The modulation wave vectors connecting different points of
the Fermi surface with the same energy gap are shown in
the absence of a supercurrent.

Fig. 4: The FCLDOS at $\omega = -44$ meV, $-16$ meV, 0 meV, 16 meV
and 44 meV (from top to bottom) and $q=0, q_c(0)$ and
$q_c(\frac{\pi}{4})$ (from left to right) in the first
Brillouin zone for a unitary defect. 

Fig. 5: The FCLDOS along the antinodal and nodal directions
at $\omega = 0$ meV and different
supercurrents for a unitary defect.

Fig. 6: The predicted, blocking-model-corrected LDOS 
$\rho_{\rm expt}({\bf R},\omega)$ at the sites $(0,0), (0,-1),
(-1,-1)$ and $(2,0)$ (from top to bottom) for supercurrents
along the antinodal (left) and nodal (right) directions, when a
unitary impurity is located at the $(0,0)$ site. Solid: $q=0$,
dash: $q=0.2\Delta^0$ and dot: $q=q_c(\phi)$. 
 
\end{document}